\documentclass[12pt]{article}
\usepackage{indentfirst} 
\usepackage{amsfonts,amssymb,amsmath}
\usepackage{bm}          
\usepackage{cite}
\usepackage{url}
\usepackage{enumitem}
\usepackage[usenames,dvipsnames]{xcolor}
\usepackage{todonotes}
\usepackage{multicol}
\usepackage{MnSymbol}
\usepackage{hyperref}
\usepackage{tikz-cd}
\usepackage[titletoc,title]{appendix}

\usetikzlibrary{arrows}

\AtBeginDocument{}%

\newcommand{\corurl}{red}
\newcommand{\corcite}{ForestGreen}
\newcommand{\corlink}{blue}

\hypersetup{linktocpage,colorlinks,urlcolor=\corurl,citecolor=\corcite,linkcolor=\corlink,
pdftitle={Hamiltonian description of the parametrized scalar field in bounded spatial regions},
pdfauthor={J.F. Barbero, J. Margalef-Bentabol, E.J.S. Villaseñor}}

\numberwithin{equation}{section}  
\begin{document}


\bibliographystyle{plain}

\title{Hamiltonian description of the parametrized scalar field in bounded spatial regions}

\author{
  {\small J. Fernando Barbero G.${}^{1,3}$, Juan Margalef-Bentabol${}^{1,2}$, and
          Eduardo J.S. Villase\~nor${}^{2,3}$} \\[4mm]
  {\small\it ${}^1$Instituto de Estructura de la Materia, CSIC} \\[-0.2cm]
  {\small\it Serrano 123, 28006 Madrid, Spain}         \\[1mm]
  {\small\it ${}^2$Grupo de Modelizaci\'on, Simulaci\'on Num\'erica
                   y Matem\'atica Industrial}  \\[-0.2cm]
  {\small\it Universidad Carlos III de Madrid} \\[-0.2cm]
  {\small\it Avda.\  de la Universidad 30, 28911 Legan\'es, Spain}            \\[1mm]
  {\small\it ${}^3$Grupo de Teor\'{\i}as de Campos y F\'{\i}sica
             Estad\'{\i}stica}\\[-2mm]
  {\small\it Instituto Gregorio Mill\'an, Universidad Carlos III de
             Madrid}\\[-2mm]
  {\small\it Unidad Asociada al Instituto de Estructura de la Materia, CSIC}
             \\[-2mm]
  {\small\it Madrid, Spain}           \\[-2mm]
  {\protect\makebox[5in]{\quad}}  
  \\
}
\date{July 17, 2015}
\maketitle
\thispagestyle{empty}   

\begin{abstract}
We study the Hamiltonian formulation for a parametrized scalar field in a regular bounded spatial region subject to Dirichlet, Neumann and Robin boundary conditions. We generalize the work carried out by a number of authors on parametrized field systems to the interesting case where spatial boundaries are present. The configuration space of our models contains both smooth scalar fields defined on the spatial manifold and spacelike embeddings from the spatial manifold to a target spacetime endowed with a fixed Lorentzian background metric. We pay particular attention to the geometry of the infinite dimensional manifold of embeddings and the description of the relevant geometric objects: the symplectic form on the primary constraint submanifold and the Hamiltonian vector fields defined on it.
\end{abstract}


\clearpage

%
%
\section{Introduction}{\label{sec_intro}}

Parametrized theories are interesting examples of diff-invariant field systems.\ They were introduced by Dirac \cite{Dirac1} circa 1950 in order to explore the idea of forgoing the use of fixed spacetime foliations with flat spatial slices, and consider a more general approach to the study of relativistic field theories. They were further analyzed by other authors, in particular Kucha\v{r}, Isham, H\'aj\'{\i}\v{c}ek and Torre \cite{Kuchar3,Hajicek,Isham1,Torre}, in the case where the spatial manifolds were taken to be closed (i.e.\ compact without boundary). Parametrized scalar fields have been recently used to explore the resolution of a number of technical issues related to the treatment of diffeomorphisms in loop quantum gravity \cite{LV1,LV2} and in the search for boundary observables \cite{Andrade}. Moreover, some interesting gravitational models, such as Einstein-Rosen waves in vacuum or coupled to massless scalar fields, are known to be dynamically equivalent to a parametrized massless scalar field with cylindrical symmetry \cite{Kuchar,BGV} (see also \cite{LR} and references therein).

The main idea in the parametrized framework is to adjoin space diffeomorphisms as dynamical variables to the usual set of fields. This is done in such a way that the solutions to the standard field equations can be naturally mapped to the one of the parametrized model and viceversa. However it is important to point out that the interpretation of parametrized models differs in some important aspects from the standard ones from which they are derived. One reason for this is the presence of gauge symmetries associated with the reparametrization invariance introduced by the diffeomorphisms.

The standard Hamiltonian formalism of mechanics is essentially geometric: dynamics takes place in the cotangent bundle associated with a space of physical configurations, a differentiable manifold itself, which is infinite-dimensional in the case of field theories. It is then necessary to be aware of the difficulties that this dimensionality issue may present and choose the right mathematical tools to deal with them. As we will show in the paper, this is not difficult once the proper formalism has been identified. A large part of the physical literature on parametrized field theories ---certainly full of useful insights at the geometric level \cite{Kuchar1,Kuchar2}--- has some technical/practical disadvantages: the first one is the local character of the standard approach (use of particular coordinate systems and a certain forgetfulness about global issues), the second is the use of distributional objects at a basic level (i.e.\ the definition of the fundamental Poisson brackets). Although distributions can, of course, be dealt with and profitably used in a completely rigorous way, they may pose some problems in practice, in particular for the type of systems that we want to study where manifolds of different dimensions (spatial manifolds and their boundaries) coexist in a non-trivial way. In this paper we will hence use coordinate-free, global, geometric methods. As we hope to show, once the proper set up is identified, computations are straightforward.

A very important feature of parametrized models is their singular character: their Lagrangians do not lead to a one-to-one fibre derivative (used to define the Legendre transformation). This means that it is not straightforward to obtain a consistent Hamiltonian description for them. The most popular approach to this problem relies on the Dirac theory of constraints \cite{Dirac2}. A more geometric way to do it uses the so called GNH method developed by Gotay, Nester and Hinds \cite{GN,Gotay,GNH}. Among the strong points of the latter are its global character and the possibility of naturally taking into account the necessary functional analytic issues relevant for field theories. In the particular case of parametrized models this method is very simple to use and, hence, will be employed here.

 The introduction of boundaries is interesting from a physical point of view. For obvious reasons, they are relevant in condensed matter systems but also in other contexts, specially in general relativity, where boundaries play a very significant role (spatial infinity, black hole horizons, holography, etc.).  A \textit{covariant} Hamiltonian approach to general relativity coupled to different types of matter fields  ---in particular, scalar fields--- in bounded spatial regions is discussed at length in \cite{Anco1,Anco2}. In the present paper we rely instead on the \textit{standard} Hamiltonian description in phase space to study parametrized scalar fields. This is the simplest parametrized field system although some more complicated models can be considered (see, for instance, \cite{Nos2}). Our results can be compared with those of \cite{Anco1,Anco2}---and checked to be compatible---by considering fixed foliations and non-dynamical metrics respectively. Notice, however, that our description uses the canonical phase space and, hence, the conditions involving momenta are not obvious in the covariant setting. An interesting application of the results that we give in the paper is the extension of the polymeric quantization carried out when no boundaries are present to these richer models with a more complicated phase space description \cite{LV1,LV2}.

To the best of our knowledge, reference \cite{Andrade} is the first paper which considers the Hamiltonian formalism for parametrized scalar fields subject to different kinds of boundary conditions (in particular of the Robin type). In that work the Dirichlet and Neumann cases are dealt with by resorting to standard Hamiltonian methods. On the other hand the Robin case is studied by using covariant Hamiltonian techniques because the boundary term introduced in the relevant action gives rise to some difficulties when going to phase space. As we show in the present paper the geometric and global methods that we use here can easily handle the Robin case while providing a very natural and simple interpretation for the results corresponding to Dirichlet and Neumann boundary conditions. We would like to mention here that our point of view is useful to understand in a clear way the bifurcation phenomenon in the constraint algorithm characteristic of parametrized gauge systems discussed in \cite{Nos2}.

The structure of the paper is the following. After this introduction we discuss in section \ref{sec_ParametrizedScalar} the action principle for the parametrized scalar field with the boundary conditions considered in the paper: Dirichlet and Robin (of which Neumann is a subcase). Section \ref{section_3+1} is devoted to the $3+1$ decomposition of the objects that we use throughout the paper. The most relevant results related to the configuration space and its tangent and cotangent bundles, where embeddings play a central role, are discussed in section \ref{sec_tangent and contangent}. Section \ref{sec_Hamiltonian formulation} is devoted to the Hamiltonian formulation. We end the paper in section \ref{sec_conclusions} with a discussion of its main results and an appendix where we collect the variations of a number of geometric objects defined on the manifold of embeddings.

%
%
\section{The parametrized scalar field in bounded spatial domains}{\label{sec_ParametrizedScalar}}

Let us consider a four dimensional Lorentzian manifold $(M,g)$ diffeomorphic to a product manifold $[t_1,t_2]\times\Sigma$, where $\Sigma$ is a smooth, orientable, compact, 3-manifold with smooth boundary (possibly empty). In this situation $\partial M$ can be written as the union $\partial_1 M\cup \partial_\Sigma M \cup \partial_2 M$ where $\partial_\Sigma M$ is diffeomorphic to $[t_1,t_2]\times\partial\Sigma$ and $\partial_i M$ ($i=1,2$) are diffeomorphic to $\Sigma$. We will assume that $\partial_i M$ are spacelike and that the Lorentzian metric $g$, with $(\varepsilon, +,+,+)$ signature, induces a Lorentzian metric $g_\partial:=j_\partial^*g$ on $\partial_\Sigma M$, where $j^*_\partial$ denotes the pullback under the inclusion map $j_{\partial}:\partial_\Sigma M\hookrightarrow M$. We have introduced the parameter $\varepsilon =-1$ to allow for a straightforward extension of our results to the Euclidean (Riemannian) case. We will restrict ourselves to the case where $(M,g)$ is time oriented.

The starting point to arrive at the action for the parametrized scalar field that we study in the paper is
\begin{equation}
S^0(\varphi):=\frac{1}{2}\int_M\left(\varepsilon g^{-1}(\mathrm{d}\varphi,\mathrm{d}\varphi)-m^2\varphi^2 \right)\mathrm{vol}_{g}-\frac{1}{2}\int_{\partial_\Sigma M}B^2\varphi^2 \mathrm{vol}_{g_\partial}\,,
\label{Action-non-parametrized}
\end{equation}
where $\varphi:M\rightarrow\mathbb{R}$ is a real scalar field on $M$, $B:\partial_\Sigma M\rightarrow\mathbb{R}$ is a fixed smooth function and we are using the metric volumes on $M$ and $\partial_\Sigma M$.  This action is the obvious generalization of the one used in \cite{Nos1} to discuss the Hamiltonian formulation for the scalar field with Robin boundary conditions (it can be essentially found in \cite[page 227]{Brezis}). Our results can be trivially extended to the case where a potential term $V(\varphi)$ is included as in \cite{Hajicek}.

The different types of boundary conditions that we consider in the paper are:
\begin{enumerate}
\item If  $\Sigma$ is closed then we have the simple example of a massive (Klein-Gordon) scalar field without boundary.
\item If $\Sigma$ has a non-empty boundary and we demand $\varphi\circ j_{\partial}=0$ we describe a massive scalar field subject to homogeneous Dirichlet boundary conditions.
\end{enumerate}
In these first two cases $B$ plays no role and can be taken to be zero.
\begin{enumerate}[resume]
\item If $\Sigma$ has a non-empty boundary and we take $B=0$ then we have a massive scalar field with homogeneous Neumann boundary conditions.
\item If $\Sigma$ has a non-empty boundary and we take $B\neq0$ then we have a massive scalar field subject to homogeneous Robin boundary conditions.
\end{enumerate}
In the last two cases no conditions on $\varphi\circ j_{\partial}$ are imposed \textit{a priori}. Notice, also, that the Robin conditions include the Neumann ones as a particular example.

Several comments are in order now. First, just mention that the non-homogeneous case is straightforward once the homogeneous one is understood (see \cite{Nos1}), so we will only consider the latter. A second comment refers to the way the variational principle works. We get the field equations by demanding that the action is stationary under admissible variations of the fields with their values kept fixed at $\partial_1M$ and $\partial_2M$.  The variations of the action are directional derivatives which may have both interior and boundary contributions. The presence of the latter does by no means imply that the action ceases to be differentiable. In fact, boundary contributions to the variations are actually the origin of \textit{natural} boundary conditions such as the Neumann or Robin ones.

Dirichlet and Robin conditions are implemented at the action level in different ways. In the first case one has to restrict oneself to field configurations (and, hence, also variations) that vanish at the boundary $\partial_\Sigma M$, whereas in the second the boundary conditions themselves appear as a consequence of the requirement that the action is stationary under arbitrary variations. The Hamiltonian description of these models based on the use of the GNH method can be found in \cite{Nos1}.

In order to parametrize the action \eqref{Action-non-parametrized} we introduce diffeomorphisms $Z:I\times\Sigma\rightarrow M$ such that, for every $t\in I:=[t_1,t_2]$,  the image $Z(\Sigma_t)$ of the slice $\Sigma_t:=\{t\}\times\Sigma$ is spacelike and $TZ.\partial_t$ is timelike and future pointing. These diffeomorphisms are taken as dynamical variables in addition to the scalar field. We consider now the action
\begin{equation}
S(\psi,Z):=\frac{1}{2}\int_{I\times\Sigma}\!\left(\varepsilon  (Z^*g)^{-1}(\mathrm{d}\psi,\mathrm{d}\psi)-m^2\psi^2 \right)\mathrm{vol}_{Z^*g}-\frac{1}{2}\int_{I\times\partial\Sigma}\!(\!B^2\circ Z)  \psi^2\mathrm{vol}_{ Z^*g_\partial}\,,
\label{Action-parametrized}
\end{equation}
where $Z^* g$ denotes the pullback of $g$ to $I\times\Sigma$ and $\psi:I\times\Sigma\rightarrow\mathbb{R}$ is a new scalar field.

Regardless of the considered boundary conditions, the evolution of the scalar field $\psi$ is given by the Klein-Gordon equation
\begin{equation}
(\varepsilon \Box_{Z^*g} +m^2)\psi=0\,,
\label{KG_eq}
\end{equation}
where the d'Alembert operator is defined in terms of the Levi-Civita connection associated with $Z^*g$. There are no extra conditions on the fields if $\Sigma$ is closed. In the case of imposing Dirichlet boundary conditions we must require that $\psi$ vanishes at $I\times\partial\Sigma$, and in the Robin (Neumann, $B=0$) case the variations of the actions provide the equations at the boundary
\[
\Big(\varepsilon  TZ^{-1}.\mathcal{V}(\psi)+(B^2\circ Z) \psi\Big)=0 \qquad\text{at }I\times\partial\Sigma
\]
where $TZ^{-1}.\mathcal{V}$ is the push-forward by $Z^{-1}$ of the unit, spacelike, outer normal $\mathcal{V}$ to $\partial_\Sigma M$. No independent field equations are obtained by varying in $Z$ as in the case when no boundaries are present, however it should be noted that the diffeomorphism $Z$ must take $I\times\partial\Sigma$ to $\partial_\Sigma M$.

%
%
\section{Some remarks on the 3+1 decomposition}\label{section_3+1}

We give here some results that we use to decompose geometric objects in $I\times\Sigma$ and get the Lagrangian for our system. To this end we introduce a unique decomposition of any tensor field on $I\times\Sigma$ with the help of the projector $\Pi:=\mathrm{Id}-\mathrm{d}t\otimes\partial_t$, which is a $(1,1)$-tensor field. For instance, vector fields $Y\in\mathfrak{X}(I\times\Sigma)$ can be written as $Y=\mathrm{d}t(Y)\partial_t+\Pi(Y)$ and one forms $\alpha\in \Omega^1(I\times\Sigma)$ as $\alpha=\alpha(\partial_t)\mathrm{d}t+\Pi(\alpha)$. We will also use the notations $\Pi_*:\mathfrak{X}(I\times\Sigma)\rightarrow\mathfrak{X}(I\times\Sigma)$ and $\Pi^*:\Omega^1(I\times\Sigma)\rightarrow\Omega^1(I\times\Sigma)$ to refer to these maps.

\subsection*{Metrics}

Any metric $\mathrm{g}$ on $I\times\Sigma$ (in particular the pullback $\mathrm{g}=Z^*g$ that we use throughout the paper) can be decomposed with the help of the identity $\mathrm{Id}=\mathrm{d}t\otimes\partial_t+\Pi$:
\begin{equation}
\mathrm{g}=\mathrm{g}(\partial_t,\partial_t)\mathrm{d}t\otimes \mathrm{d}t+\mathrm{d}t\otimes \mathrm{g}(\partial_t,\Pi_*\cdot)+\mathrm{g}(\partial_t,\Pi_*\cdot)\otimes \mathrm{d}t+\mathrm{g}(\Pi_*\cdot,\Pi_*\cdot)\,.\label{3+1metric}
\end{equation}
This is usually written in the form
\begin{equation}
\mathrm{g}=\big(\varepsilon \tilde{N}^2+\tilde{\gamma}^{-1}(\tilde{\beta},\tilde{\beta})\big)\mathrm{d}t\otimes \mathrm{d}t+\mathrm{d}t\otimes \tilde{\beta}+\tilde{\beta} \otimes \mathrm{d}t+\tilde{\gamma}\,,
\label{3+1metric2}
\end{equation}
where the so called \textit{shift} $\tilde{\beta}\in\Omega^1(I\times\Sigma)$ satisfies $\tilde{\beta}(\partial_t)=0$, $\tilde{\gamma}$ is a symmetric $(2,0)$-tensor satisfying $\tilde{\gamma}(\cdot,\partial_t)=0$ and $\tilde{\gamma}^{-1}$ (a symmetric $(0,2)$-tensor) is the unique pseudo-inverse of $\tilde{\gamma}$ such that $\tilde{\gamma}^{-1}\cdot \tilde{\gamma} =\Pi$ and $\tilde{\gamma}^{-1}(\cdot,\mathrm{d}t)=0$. Finally the nowhere vanishing function $\tilde{N}\in C^\infty(I\times\Sigma)$ is the \textit{lapse}. The inverse metric $\mathrm{g}^{-1}$ can be conveniently written as
\begin{equation}
\mathrm{g}^{-1}=\frac{\varepsilon }{\tilde{N}^2}\big(\partial_t-\tilde{\gamma}^{-1}(\tilde{\beta},\cdot)\big)\otimes\big(\partial_t-\tilde{\gamma}^{-1}(\tilde{\beta},\cdot)\big)+\tilde{\gamma}^{-1}\,.
\label{3+1inverse-metric}
\end{equation}

%
%
\subsection*{Diffeomorphisms and embeddings}

Let $Z:I\times \Sigma\rightarrow M$ be a diffeomorphism such that $Z(\Sigma_t)\subset M$ is spacelike for each $t\in I$. We can define a decomposition of $Z$ and its tangent map $TZ$ adapted to the product structure of $I\times\Sigma$ by using the projector tensor field $\Pi$ introduced above. The tangent map $TZ:T(I\times\Sigma)\rightarrow TM$ acting on a vector field $V\in\mathfrak{X}(I\times\Sigma)$ gives
\begin{equation}
TZ.V=TZ.(\mathrm{d}t(V)\partial_t+\Pi_*V)=\mathrm{d}t(V)TZ.\partial_t+TZ.\Pi_*V\,.
\label{tangentmap}
\end{equation}
We can, hence, write $TZ=\dot{Z}\mathrm{d}t+\underline{TZ}$ with $\dot{Z}:=TZ.\partial_t$ and $\underline{TZ}:=TZ.\Pi_*$. Notice that $\underline{TZ}.V\in \mathfrak{X}(M)$ is tangent to the submanifold $Z(\Sigma_t)$ for each $t\in I$.

We can apply the preceding results regarding the decompositions of the metric and diffeomorphisms to the particular case of $\mathrm{g}=Z^*g$. Let $V_1,V_2\in\mathfrak{X}(I\times\Sigma)$ then
\begin{align}
(Z^*g)(V_1,V_2)&=g(TZ.V_1,TZ.V_2)=g(\mathrm{d}t(V_1)\dot{Z}+\underline{TZ}.V_1,\mathrm{d}t(V_2)\dot{Z}+\underline{TZ}.V_2)\nonumber\\
&=g(\dot{Z},\dot{Z})\mathrm{d}t(V_1)\mathrm{d}t(V_2)+\mathrm{d}t(V_1)g(\dot{Z},\underline{TZ}.V_2)+\mathrm{d}t(V_2)g(\dot{Z},\underline{TZ}.V_1)\nonumber\\
&\phantom{=}+g(\underline{TZ}.V_1,\underline{TZ}.V_2)\,.\nonumber
\end{align}
Comparing the previous expression with \eqref{3+1metric2}, we write
\begin{eqnarray}
\tilde{\gamma}_Z&=&g(\underline{TZ}\cdot,\underline{TZ}\cdot)\,,\label{tildegamma}\\
\tilde{\beta}_Z&=&g(\dot{Z},\underline{TZ}\cdot)\,,\label{tildebeta}\\
\varepsilon  \tilde{N}_Z^2&=&g(\dot{Z},\dot{Z})-\tilde{\gamma}^{-1}(\tilde{\beta}_Z,\tilde{\beta}_Z)\,.\label{tildeN}
\end{eqnarray}
Notice that $\tilde{\gamma}_Z$ is a symmetric $(2,0)$-tensor field on $I\times\Sigma$ but \textit{it is not} a metric as $\tilde{\gamma}_Z(\partial_t,\cdot)=0$.

For each $t\in I$ let us define the map $\jmath_t:\Sigma\rightarrow I\times\Sigma:s\mapsto (t,s)$, we have then that $\jmath_t(\Sigma)=\Sigma_t$ and $Z_t:=Z\circ \jmath_t:\Sigma \rightarrow M$ is an embedding for each $t\in I$. The fact that $Z$ is a diffeomorphism of the  type considered in the paper guarantees that the images $Z_t(\Sigma)$, $t\in I$,  provide a foliation of $M$ by spacelike hypersurfaces. The diffeomorphism itself can be reinterpreted as a smooth\footnote{We will not discuss topological issues here but just mention that they can be handled by the convenient calculus approach of Michor \cite{Michor2}.} curve of embeddings. By using the $\jmath_t$ map we can pullback the tensors \eqref{tildegamma}-\eqref{tildeN}  to $\Sigma$. This way we get
\begin{eqnarray}
\gamma_{Z_t}&:=&\jmath_t^*\tilde{\gamma}_Z= Z_t^*g\,.\label{gammaZt}\\
\beta_{(Z_t,\dot{Z}_t)}&:=&\jmath_t^*\tilde{\beta}_Z= g(\dot{Z}_t,TZ_t\cdot)\,,\label{betaZt}\\
\varepsilon  N_{(Z_t,\dot{Z}_t)}^2&:=&\varepsilon \jmath_t^*\tilde{N}^2_Z= g(\dot{Z}_t,\dot{Z}_t)-\gamma^{-1}_{Z_t}(\beta_{(Z_t,\dot{Z}_t)},\beta_{(Z_t,\dot{Z}_t)})\,.\label{NZt}
\end{eqnarray}
Although in the next section we will discuss with more detail the manifold of embeddings, it is important to highlight at this point that the Riemannian metric $\gamma_{Z_t}$ depends only on $Z_t$ whereas both $\beta_{(Z_t,\dot{Z}_t)}$ and $\varepsilon  N_{(Z_t,\dot{Z}_t)}^2$ depend also on the velocity $\dot{Z}_t$, which is a vector field in $M$ along the embedding $Z_t$. Moreover, $\beta_{(Z_t,\dot{Z}_t)}$ is linear in $\dot{Z}_t$ as a consequence of \eqref{betaZt}, as also is $N_{(Z_t,\dot{Z}_t)}$, because
\begin{equation}
N_{(Z_t,\dot{Z}_t)}=\varepsilon n_{Z_t}(\dot{Z}_t):=\varepsilon g(n_{Z_t},\dot{Z}_t)\,,
\label{nX}
\end{equation}
where we denote $n_{Z_t}$ the future-pointing unit $g$-normal to $Z_t(\Sigma)\subset M$. In a similar fashion we introduce $\theta_{Z_t}$ as the future-pointing unit $g_\partial$-normal to $Z_t(\partial\Sigma)\subset \partial_\Sigma M$.

We introduce now two objects that will be useful in the following. If $X:\Sigma\hookrightarrow M$ is an embedding we define
\begin{eqnarray}
&&(\tau_X)^\alpha_a:=(TX)^\alpha_a\,,\label{tauX}\\
&&(e_X)^a_\alpha:=g_{\alpha\beta}(TX)^\beta_b\gamma_X^{ab}\,.\label{eX}
\end{eqnarray}
Here and in the following we will use the Penrose abstract index notation when convenient ($\alpha$, $\beta$, etc.\ are abstract indices on $M$ and $a$, $b$, etc.\ abstract indices on $\Sigma$). These objects satisfy
\begin{equation}
(\tau_X)^\alpha_a(e_X)^b_\alpha=\delta_a^b\,,\quad (\tau_X)^\alpha_a(e_X)^a_\beta=\delta^\alpha_\beta - \varepsilon n_X^\alpha n_{X\,\beta}\,,
\end{equation}
where $n_\alpha=g_{\alpha\beta} n^\beta$. From equation \eqref{betaZt} we can write
\begin{equation}
\beta^a_{(Z_t,\dot{Z}_t)}=(e_{Z_t})^a_\mu\dot{Z}_t^\mu\,.
\label{betasupera}
\end{equation}

It is straightforward to write now \eqref{Action-parametrized} in the form
\begin{align}
S(\psi,Z)&=\frac{1}{2}\int_I \mathrm{d}t\!\int_{\Sigma}\varepsilon n_{Z_t}(\dot{Z}_t) \left(\left[\frac{\dot{\psi}_t-(d\psi_t)_b(e_{Z_t})^b_\beta\dot{Z}_t^\beta}{n_{Z_t}(\dot{Z}_t)}\right]^2\!\!+\varepsilon \gamma^{-1}_{Z_t}(d\psi_t,d\psi_t)-m^2 \psi^2_t\right)\mathrm{vol}_{\gamma_{Z_t}}
\nonumber\\
&\phantom{=}-\frac{1}{2}\int_I \mathrm{d}t\!\int_{\partial\Sigma} \varepsilon \theta_{Z_t}(\dot{Z_t})b_{Z_t}^2 \psi_t ^2\mathrm{vol}_{\gamma_{\partial Z_t}}\label{action3+1}
\end{align}
where $\psi_t:=\jmath_t^*\psi$, $\dot\psi_t:=\jmath_t^*\dot\psi$, $\dot{\psi}:=\mathrm{d}\psi(\partial_t)$, $b_{Z_t}:=B\circ Z_t\circ\imath_\partial$, $\gamma_{\partial Z_t}:= \imath_\partial^*\gamma_{Z_t}$ and $\imath_\partial:\partial\Sigma\hookrightarrow\Sigma$. Although it is possible to read off the formal expression of the Lagrangian for our model from the previous equation, it is important to pay attention to its domain, an appropriate subset of the tangent bundle of the configuration space. We devote the next section to this issue.
%
%
\section{Dynamics: geometric arena}\label{sec_tangent and contangent}
\subsection*{Configuration Space}
The configuration spaces of the different parametrized scalar fields systems that we consider in the following sections are
\begin{align}
\mathcal{C}_D&:=C^\infty_D(\Sigma)\times \mathrm{Emb}^\partial_{g\textrm{-sl}}(\Sigma,M)\subset C^\infty(\Sigma)\times \mathrm{Emb}_{g\textrm{-sl}}(\Sigma,M)\,,\\
\mathcal{C}_R&:=C^\infty_R(\Sigma)\times \mathrm{Emb}^\partial_{g\textrm{-sl}}(\Sigma,M)\subset C^\infty(\Sigma)\times \mathrm{Emb}_{g\textrm{-sl}}(\Sigma,M)\,.
\end{align}
Here $C^\infty_R(\Sigma)=C^\infty(\Sigma)$ consists of smooth scalar fields while the elements of $C^\infty_D(\Sigma)$ are smooth functions that vanish at the boundary $\partial\Sigma$.  $\mathrm{Emb}_{g\textrm{-sl}}(\Sigma,M)$ is the space of smooth, spacelike embeddings of $\Sigma$ in $(M,g)$. This space has been studied in \cite{Isham1,Hajicek} (see also \cite{Bauer}). Our space $\mathrm{Emb}^\partial_{g\textrm{-sl}}(\Sigma,M)$ is the subclass of such embeddings taking $\partial\Sigma$ to $\partial_\Sigma M$. For $J\in\{D,R\}$ we will denote a configuration $(q,X)\in \mathcal{C}_J$ where $q:\Sigma\rightarrow\mathbb{R}$ is a smooth scalar field and $X:\Sigma\hookrightarrow M$ is an embedding.

\subsection*{Velocity Space}

In the following we will use the standard notation for fibre bundles so, for instance, $\Gamma(X^*TM)$ will denote sections of the pullback bundle $X^*TM$ defined by the embedding $X$. The velocity phase-space $T\mathcal{C}_J$ is the product manifold
\begin{equation}
T\mathcal{C}_J=TC^\infty_J(\Sigma)\times T\mathrm{Emb}^\partial_{g\textrm{-sl}}(\Sigma,M)\,,
\end{equation}
 where $TC^\infty_J(\Sigma)=C^\infty_J(\Sigma)\times C^\infty_J(\Sigma)$ and  for each $X\in\mathrm{Emb}^\partial_{g\textrm{-sl}}(\Sigma,M)$ we have:
 \[T_X\mathrm{Emb}^\partial_{g\textrm{-sl}}(\Sigma,M) = \Gamma^\partial(X^*TM):=\{V_X\in \Gamma(X^*TM)\,:\, V_X|\partial \Sigma \in \Gamma(X^*T\partial_\Sigma M) \}.\]
This can be restated in terms of the commutativity of the following diagrams

\begin{center}
\begin{tikzcd}
  & &  TM\arrow{d}\\
   \Sigma \arrow{rru}{V_X} \arrow[swap,hook]{rr}{X} &  & M
\end{tikzcd}
\qquad\qquad
\begin{tikzcd}
 &  & T\partial_\Sigma M\arrow{d}\\
     \partial\Sigma \arrow{rru}{V_X|\partial\Sigma} \arrow[swap,hook]{rr}{X|\partial\Sigma}  &  & \partial_\Sigma M
\end{tikzcd}
\end{center}
where the vertical arrows represent the natural projections in the respective tangent bundles. The condition defined by the second diagram implies
\begin{equation}
g(V_X|\partial\Sigma, \nu_X)=0\,,
\label{nuperp}
\end{equation}
where $\nu_X=\mathcal{V}\circ X\circ\imath_\partial$ and $\mathcal{V}$ is the unit, spacelike, outer normal to $\partial_\Sigma M$. For generic elements of $T_X\mathrm{Emb}_{g\textrm{-sl}}(\Sigma,M)=\Gamma(X^*TM)$ only the left diagram applies, as is the case for $n_X$. In the following, a typical element of the velocity phase-space $T_{(q,X)}\mathcal{C}_J$ will be denoted as $\mathrm{v}_{(q,X)}=(v,V_X)$, where $v\in C^\infty_J(\Sigma)$ and $V_X\in \Gamma^\partial(X^*TM)$.

A comment about the notation that we have used so far and will be used throughout the paper is in order now: if we consider some tensor field $V\in\Gamma(T^{r,s}\textrm{Emb}(\Sigma,M))$ over the space of embeddings we will denote $V_X\in T_X^{r,s}\textrm{Emb}(\Sigma,M)$. The same subindex notation will be applied for more general objects like $\tau$ and $e$, that whenever they are considered over a fixed embedding $X$, will be denoted $\tau_X,e_X$. Therefore if we consider $V\in\mathfrak{X}(\textrm{Emb}(\Sigma,M))$ then $V_X\in T_X\mathrm{Emb}_{g\textrm{-sl}}(\Sigma,M)$, and we have just seen that this element can be considered as a vector field over the embedding $X$. It is clear then that such vector field can be decomposed in the form $V_X=V_X^\perp n_X + \tau_X.V_X^\top$, where $\varepsilon  V_X^\perp:=g(V_X,n_X)\in C^\infty_J(\Sigma)$ and  $V_X^\top\in \mathfrak{X}(\Sigma)$ is defined by $\tau_X.V_X^\top:=V_X-V_X^\perp n_X$. Such decomposition can be made over the space of embeddings writing simply
\[
V=V^\perp n + \tau.V^\top\in\mathfrak{X}(\textrm{Emb}(\Sigma,M)).
\]
Altough here both addens are vector fields on the space of embedding, notice that $V^\perp:\textrm{Emb}(\Sigma,M)\to C^\infty(\Sigma)$ and $V^\top:\textrm{Emb}(\Sigma,M)\to\mathfrak{X}(\Sigma)$.

Finally notice that if $M=I\times \mathbb{R}^3$, the tangent bundle $T\mathrm{Emb}_{g\textrm{-sl}}(\Sigma,M)$ is the trivial bundle $T\mathrm{Emb}_{g\textrm{-sl}}(\Sigma,M) =\mathrm{Emb}_{g\textrm{-sl}}(\Sigma,M)\times C^\infty(\Sigma,M)$.

\subsection*{Phase Space}
The phase-space $T^*\mathcal{C}_J$ is the product manifold
\begin{equation}
T^*\mathcal{C}_J=T^*C^\infty_J(\Sigma)\times T^*\mathrm{Emb}^\partial_{g\textrm{-sl}}(\Sigma,M)\,,
\end{equation}
 where $T^*C^\infty_J(\Sigma)=C^\infty_J(\Sigma)\times C^\infty_J(\Sigma)'$ and
\begin{equation}
T^*_X\mathrm{Emb}^\partial_{g\textrm{-sl}}(\Sigma,M) =\{\bm{P}_X\,|\, \bm{P}_X:\Gamma^\partial(X^*TM)\rightarrow\mathbb{R}\,\textrm{ linear and continuous}\}\,.
\end{equation}
In the case $M=I\times\mathbb{R}^3$ we have that $T^*\mathrm{Emb}_{g\textrm{-sl}}(\Sigma,M)$ is a trivial bundle over the base $\mathrm{Emb}_{g\textrm{-sl}}(\Sigma,M)$. A typical point of $T_{(q,X)}^*\mathcal{C}_J$ is of the form $\bm{\mathrm{p}}_{(q,X)}=(\bm{p},\bm{P}_X)$ where  $\bm{p}\in C^\infty_J(\Sigma)'$ is a distribution, i.e.\ a continuous linear functional $\bm{p}: C^\infty_J(\Sigma)\rightarrow \mathbb{R}$,  and $\bm{P}_X:\Gamma^\partial(X^*TM)\rightarrow\mathbb{R}$ is a continuous linear functional defined on the space of vector fields along the embedding $X$. The phase-space $T^*\mathcal{C}_J$ is equipped with the symplectic form $\Omega$ defined by
\begin{equation}
\Omega_{\bm{\mathrm{p}}_{(q,X)}}(Y_1,Y_2)
=\bm{Y}_{\textsf{p}\,2}\left(Y_{\textsf{q}\,1}\right)-\bm{Y}_{\textsf{p}\,1}\left(Y_{\textsf{q}\,2}\right)+\bm{Y}_{\textsf{P}\,2}\left(Y_{\textsf{X}\,1}\right)-\bm{Y}_{\textsf{P}\,1}\left(Y_{\textsf{X}\,2}\right)\,,
\end{equation}
where
\begin{eqnarray*}
 Y:=\Big(((q,X),(\bm{p},\bm{P}_X)),((Y_{\textsf{q}\,i},Y_{\textsf{X}\,i}),(\bm{Y}_{\bm{\textsf{p}}\,i},\bm{Y}_{\bm{\textsf{P}}\,i}))\Big)\in T_{\bm{\mathrm{p}}_{(q,X)}} T^*\mathcal{C}_J\,.
 \end{eqnarray*}
As $Y_i$ are  tangent vectors of the phase space, we have $Y_{\textsf{q}\,i}\in C^\infty_J(\Sigma)$, $Y_{\textsf{X}\,i}\in \Gamma^\partial(X^*TM)$, $\bm{Y}_{\bm{\textsf{p}}\,i}\in C^\infty_J(\Sigma)'$ and $\bm{Y}_{\bm{\textsf{P}}\,i}:\Gamma^\partial(X^*TM)\rightarrow\mathbb{R}$. Notice that the sans-serif  subindices are used to denote the components of the vector field, in particular the subindex $\textsf{X}$ of $Y_{\textsf{X}}$ is entirely different from the subindex $X$ used, for instance, in $n_X$, where it denotes the evaluation of the vector field $n\in\mathfrak{X}(\textrm{Emb}(\Sigma,M))$ at the point $X$.

As we will see,  the distributions $\bm{p}$ and $\bm{P}_X$ that we will need in the paper can be defined in terms of a scalar field  $p\in C^\infty(\Sigma)$ and two covector fields along $X$; $P_X:\Gamma^\partial(X^*TM)\rightarrow C^\infty(\Sigma)$ and $P_{\partial X}:\Gamma^\partial(X^*T\partial_\Sigma M)\rightarrow C^\infty(\partial\Sigma)$. They have the form
\begin{align}
\bm{p}(v)&=\int_\Sigma p v\,\mathrm{vol}_{\Sigma}\,,
\label{eq:momentosQ}\\
\bm{P}_X(V)&=  \int_\Sigma   P_X(V)\,\mathrm{vol}_{\Sigma}+\int_{\partial\Sigma} P_{\partial X} (V)\,\mathrm{vol}_{\partial \Sigma}\,.\label{eq:momentosX}
\end{align}
Here  $\mathrm{vol}_{\Sigma}$ and $\mathrm{vol}_{\partial \Sigma}$ are fixed  volume forms  on $\Sigma$ and $\partial \Sigma$,  respectively.  These distributions can be alternatively  expressed in terms of the metric volume elements of the Riemannian metrics $\gamma_X= X^*g$ and $\gamma_{\partial X}:=(X\circ \imath_\partial)^*g= \imath_\partial^*\gamma_X$:
\begin{align}
\bm{p}(v)&=\int_\Sigma p v\,\mathrm{vol}_{\Sigma}=\int_\Sigma \frac{p v}{\sqrt{\gamma_X}}\,\mathrm{vol}_{\gamma_X}\,,\nonumber
\\
\bm{P}_X(V)&=\int_\Sigma   P_X(V)\,\mathrm{vol}_{\Sigma}+\!\int_{\partial\Sigma} P_{\partial X} (V)\,\mathrm{vol}_{\partial \Sigma}
=\int_\Sigma   \frac{P_X(V)}{\sqrt{\gamma_X}}\,\mathrm{vol}_{\gamma_X}+\!\int_{\partial\Sigma} \frac{P_{\partial X}(V)}{\sqrt{\gamma_{\partial X}}} \,\mathrm{vol}_{\gamma_{\partial X}}\,.\nonumber
\end{align}
After having introduced these geometric elements we discuss now the Hamiltonian formulation.

%
%
\section{Hamiltonian formulation}\label{sec_Hamiltonian formulation}

From the $3+1$ expression \eqref{action3+1} of the action, we can define the Lagrangians $L_J:\mathcal{D}_J\to\mathbb{R}$ for $J\in\{D,R\}$ given by
\begin{align}
L_J(\mathrm{v}_{(q,X)})&=\frac{1}{2}\int_\Sigma \!\! \varepsilon  n_{X}(V)\left( \left(\frac{v-e_{X}(\mathrm{d}q,V)}{n_X(V)}\right)^2+\varepsilon  \gamma_{X}^{-1}(\mathrm{d}q,\mathrm{d}q)-m^2q^2\right) \mathrm{vol}_{\gamma_X}\nonumber\\
&\phantom{=}-\frac{1}{2}\int_{\partial \Sigma} \varepsilon \theta_X(V)b_X^2q ^2\mathrm{vol}_{\gamma_{\partial X}}\,.\label{Lagrangian}
\end{align}
where
\begin{align*}
  \mathcal{D}_J&:=\left\{\mathrm{v}_{(q,X)}\in T\mathcal{C}_J\ :\ \varepsilon n_X(V)>0\right\}\,.
\end{align*}

The Hamiltonian formulation of dynamics is defined in the cotangent bundle of the configuration space. In order to go from the tangent to the cotangent bundle we need to compute the fibre derivative defined by our Lagrangian. This is simply
\begin{align}
FL_J(\mathrm{v}_{(q,X)})\cdot \overline{\mathrm{v}}_{(q,X)} &= -\int_\Sigma\left(\!\left(\frac{v\!-\!e_{X}(\mathrm{d}q,\!V)}{\varepsilon  \,n_X(V)}\right)
e_X(\mathrm{d}q,\!\overline{V})\right)\mathrm{vol}_{\gamma_X}\label{fibre}\\
&+
\frac{1}{2}\int_\Sigma\left(\left( \varepsilon\gamma_{X}^{-1}(\mathrm{d}q,\!\mathrm{d}q)-  m^2q^2-   \!\left(\frac{v\!
-\!e_{X}(\mathrm{d}q,\!V)}{n_X(V)}\right)^2             \right)\varepsilon n_X(\overline{V})\right)\mathrm{vol}_{\gamma_X} \nonumber\\
&-\frac{1}{2}\int_{\partial \Sigma} \varepsilon\theta_X(\overline{V})b_X^2q^2\mathrm{vol}_{\gamma_{\partial X}}\nonumber\\
&+\int_\Sigma\frac{v-e_X(\mathrm{d}q,V)}{\varepsilon \, n_X(V)}\overline{v}\,\mathrm{vol}_{\gamma_X}\nonumber\,.
\end{align}
The fibre derivative defines, among other things, the canonical momenta. Calling
\begin{align*}
 (P_X)_\alpha &=-\sqrt{\gamma_X}\left(\frac{v\!-\!e_{X}(\mathrm{d}q,\!V)}{\varepsilon  \,n_X(V)}\right)(e_X)^a_\alpha (\mathrm{d}q)_a\\
 &\phantom{=\ }+
\frac{\sqrt{\gamma_X}}{2} \left( \varepsilon\gamma_{X}^{-1}(\mathrm{d}q,\!\mathrm{d}q)-  m^2q^2-   \!\left(\frac{v\!
-\!e_{X}(\mathrm{d}q,\!V)}{n_X(V)}\right)^2             \right)\varepsilon (n_X)_\alpha\,,\\
(P_{\partial X})_\alpha &=-\frac{\sqrt{\gamma_{\partial X}} }{2} b_X^2q ^2  \varepsilon(\theta_X)_\alpha\,,\\
p&=\frac{v-e_X(\mathrm{d}q,V)}{\varepsilon \, n_X(V)}\sqrt{\gamma_X}\,,
\end{align*}
it can be seen that the distributions defined in \eqref{fibre} belong to the class defined by equations \eqref{eq:momentosQ}-\eqref{eq:momentosX}.
If we introduce now
\begin{eqnarray}
\mathcal{H}_a(q,p)&=&p(\mathrm{d}q)_a\,,\label{Ha}\\
\mathcal{H}_\perp(q,p,X)&=&\frac{\sqrt{\gamma_X}}{2}\left(m^2q^2+\frac{p^2}{\gamma_X}-\varepsilon \gamma^{-1}_X(\mathrm{d}q,\mathrm{d}q)\right)\,,\label{Hperpr}\\
\mathcal{H}^\partial_\perp(q,X)&=&\frac{\sqrt{\gamma_{\partial X}}}{2}b_X^2 q^2|\partial\Sigma\,,\label{Hperprbdry}
\end{eqnarray}
the primary constraint submanifold $\mathcal{M}_1^J:=FL_J(\mathcal{D}_J)$ can be written as $\mathcal{M}_1^J=\{ C_\alpha=0_\alpha , C^\partial_\alpha=0_\alpha\}$ where
\begin{eqnarray}
C_\alpha(q,X,p,P)&:= &(P_X)_\alpha + \varepsilon \,(n_X)_\alpha \mathcal{H}_\perp(q,p,X) +(e_X)^a_\alpha  \mathcal{H}_a(q,p)  \,,\label{constraint_bulk}\\\
C^\partial_\alpha(q,X,P_\partial)&:=&(P_{\partial X})_\alpha+\varepsilon (\theta_X)_\alpha \mathcal{H}^\partial_\perp(q,X)  \,.\label{constraint_bdry}\
\end{eqnarray}

As the Lagrangian is homogeneous of degree one, it is immediate that the energy
\begin{equation}
E\big(\mathrm{v}_{(q,X)}\big)=FL_J\big(\mathrm{v}_{(q,X)}\big)\cdot \mathrm{v}_{(q,X)}-L_J(\mathrm{v}_{(q,X)})
\end{equation}
vanishes in the full tangent bundle of the configuration space. The Hamiltonian $H$ is defined on $\mathcal{M}_1^J$ as $E=:H\circ FL_J$ and, hence, it is also zero.

The equation defining the Hamiltonian vector fields $Z$ on $\mathcal{M}_1^J$ is simply
\begin{equation}
i_{Z}\omega=0
\label{eqHamVfields}
\end{equation}
where $\omega$ is the pullback of $\Omega$ to $\mathcal{M}_1^J$ and $Z:=(((q,X),(\bm{p},\bm{P})),((Z_{\textsf{q}},Z_{\textsf{X}}),(\bm{Z}_{\textsf{p}},\bm{Z}_{\bm{\textsf{P}}})))$ denotes a vector field on $\mathcal{M}_1^J$. Notice that the components $(\bm{Z}_{\textsf{p}},\bm{Z}_{\bm{\textsf{P}}})$ are of the form \eqref{eq:momentosQ} and \eqref{eq:momentosX} respectively.

The easiest way to solve the equation for the Hamiltonian vector field is to find the solutions to $\omega(Y,Z)=0$ for all $Y\in \mathfrak{X}(\mathcal{M}_1^J)$. As we are looking at $\mathcal{M}_1^J$ as a submanifold embedded in phase space we must find a convenient way to characterize and use these fields. In the present case this can be done by requiring the vanishing of the directional derivatives of the constraints along the field.

Let us consider a point $\bm{\mathrm{p}}_{(q,X)}=((q,X),(p,P,P_\partial))$ in phase space, then $\omega$ is given by
\begin{align}
\omega_{\bm{\mathrm{p}}_{(q,X)}}(Y,Z)&=\int_\Sigma \frac{1}{\sqrt{\gamma_X}}(Z_{\textsf{p}}   Y_{\textsf{q}}-Y_{\textsf{p}}Z_{\textsf{q}}+Z_{{\textsf{P}}\alpha}Y_{\textsf{X}}^\alpha-Y_{{\textsf{P}}\alpha} Z_{\textsf{X}}^\alpha)\mathrm{vol}_{\gamma_X}\label{symplectic}\\
&\phantom{=}+\int_{\partial\Sigma}\frac{1}{\sqrt{\gamma_{\partial X}}}
( Z^\partial_{\textsf{P} \alpha} Y_{\textsf{X}}^\alpha - Y^\partial_{\textsf{P} \alpha} Z_{\textsf{X}}^\alpha ) \mathrm{vol}_{\gamma_{\partial X}}
\nonumber
\end{align}
where
\begin{align}
Y_{\textsf{P}\alpha}&=D\left(\rule{0ex}{2.3ex}-\varepsilon n_{\alpha}\mathcal{H}_\perp -e^b_{\alpha}\mathcal{H}_b\right)\,, &
Z_{\textsf{P}\alpha}&=D\left(\rule{0ex}{2.3ex}-\varepsilon n_{\alpha}\mathcal{H}_\perp -e^b_{\alpha}\mathcal{H}_b\right)\,,\label{Y}\\
Y^\partial_{\textsf{P}\alpha}&=D\left(\rule{0ex}{2.3ex}-\varepsilon \theta_{\alpha} \mathcal{H}^\partial_\perp \right)\,, &
Z^\partial_{\textsf{P}\alpha}&=D\left(\rule{0ex}{2.3ex}-\varepsilon \theta_{\alpha} \mathcal{H}^\partial_\perp \right) \label{y}
\end{align}
denote the variations \eqref{up} and \eqref{down} given in the appendix \ref{appendix2}. The concrete form of $\omega(Y,Z)$, obtained by a long but straightforward computation, is given by
\begin{align}\label{granecuacion1}
\omega(Y,&Z)=\\
\int_\Sigma&\left[Z_\mathsf{q}-\left(\frac{p}{\sqrt{\gamma}}\varepsilon n_\alpha+e^b_\alpha(dq)_b\right)Z_{\mathsf{X}}^\alpha\right]\left[\frac{Y_{\textsf{p}}}{\sqrt{\gamma}}+ m^2q  Y^\perp_{\textsf{X}}+\mathrm{div}_{\gamma}\left( \varepsilon Y^\perp_{\textsf{X}} \mathrm{grad}_{\gamma} q -\frac{p}{\sqrt{\gamma}}Y^\top_\textsf{X}
\right)\right]\mathrm{vol}_{\gamma}\nonumber\\
-\int_\Sigma&\left[Y_\mathsf{q}-\left(\frac{p}{\sqrt{\gamma}}\varepsilon n_\alpha+e^b_\alpha(dq)_b\right)Y_{\mathsf{X}}^\alpha\right]\left[\frac{Z_{\textsf{p}}}{\sqrt{\gamma}}+ m^2q  Z^\perp_{\textsf{X}}+\mathrm{div}_{\gamma}\left( \varepsilon Z^\perp_{\textsf{X}} \mathrm{grad}_{\gamma} q -\frac{p}{\sqrt{\gamma}}Z^\top_\textsf{X} \right)\right]\mathrm{vol}_{\gamma}\nonumber\\
+\int_{\partial\Sigma}&\left[Y_\mathsf{q}-\left(\frac{p}{\sqrt{\gamma}}\varepsilon n_\alpha+e^b_\alpha(dq)_b\right)Y_{\mathsf{X}}^\alpha\right]\varepsilon Z_\mathsf{X}^\perp\left(\frac{p}{\sqrt{\gamma}}\varepsilon n_\beta+e^a_\beta(dq)_a-\varepsilon b^2q\nu_\beta\right)\frac{\nu^\beta}{|\nu^\top|}\mathrm{vol}_{\partial\gamma}\nonumber\\
-\int_{\partial\Sigma}&\left[Z_\mathsf{q}-\left(\frac{p}{\sqrt{\gamma}}\varepsilon n_\alpha+e^b_\alpha(dq)_b\right)Z_{\mathsf{X}}^\alpha\right]\varepsilon Y_\mathsf{X}^\perp\left(\frac{p}{\sqrt{\gamma}}\varepsilon n_\beta+e^a_\beta(dq)_a-\varepsilon b^2q\nu_\beta\right)\frac{\nu^\beta}{|\nu^\top|}\mathrm{vol}_{\partial\gamma}\,,\nonumber
\end{align}
where we have written $\nu_X=\tau\cdot\nu^\top_X+\nu^\perp n_X$, $|\nu_X^\top|^2:=\gamma_X(\nu_X^\top,\nu_X^\top)$ and we have dropped the $X$ subindex in all the elements to render the last expression more compact. Notice that due to the antisymmetry of $\omega$, the Weingarten maps $K_X$ and $K_X^\partial$ appearing in equations \eqref{up} and \eqref{down}, cancel out in the final form of \eqref{granecuacion1}.

Requiring the previous equation to be zero for every $Y\in\mathfrak{X}(\mathcal{M}_1^J)$, we find that the Hamiltonian vector field in the interior of $\Sigma$ is
\begin{eqnarray}
Z_{\textsf{q}}&=&\frac{p}{\sqrt{\gamma_X}}Z_{\textsf{X}}^\perp+\mathcal{L}_{Z_{\textsf{X}}^\top}q=Z^\alpha_{\textsf{X}}\left(\frac{p}{\sqrt{\gamma_X}} \varepsilon  n_{X\, \alpha} +(e_X)^a_\alpha (\mathrm{d}q)_a \right)\label{ZQ}\\
Z_{\textsf{p}}&=&-m^2\sqrt{\gamma_X}\,q Z_{\textsf{X}}^\perp+ \sqrt{\gamma_X}\,\mathrm{div}_{\gamma_X}\left(\frac{p}{\sqrt{\gamma_X}}Z_{\textsf{X}}^\top-\varepsilon Z_{\textsf{X}}^\perp \mathrm{grad}_{\gamma_X} q\right)\label{ZP}
\end{eqnarray}
for all the types of boundary conditions that we consider here and, also,  when $\Sigma$ has no boundary. Notice that,  as we are working with smooth objects, equations (\ref{ZQ}, \ref{ZP}) must be extended by continuity to $\partial \Sigma$. In the preceding equations the gradient and divergence are defined in terms of the metric $\gamma_X$ in the standard way, $\mathcal{L}_{V^\top}$ denotes the Lie derivative along a vector field $V^\top\in\mathfrak{X}(\Sigma)$ and we have used the decomposition
\[
Z_{\textsf{X}}=Z_{\textsf{X}}^\perp n + \tau.Z_{\textsf{X}}^\top\,.
\]
When solving equation \eqref{eqHamVfields} no restriction over $Z_{\textsf{X}}$ in the interior of $\Sigma$ arises, hence it can be chosen freely (within the class of regular objects that we are considering).

Equations (\ref{ZQ}, \ref{ZP}) coincide with those appearing in the classic papers on the subject \cite{Hajicek}. For closed spatial manifolds this result is all that is needed to get the full Hamiltonian description, however, in the presence of boundaries some extra conditions may appear. In order to see this notice that, although the last boundary integral in \eqref{granecuacion1} vanishes, because \eqref{ZQ} is also true at the boundary, we need to require that
\begin{equation}\label{eq last boundary integral}
\int_{\partial\Sigma}\left[Y_{\textsf{q}}-\left(\frac{p}{\sqrt{\gamma}}\varepsilon n_{\alpha}+e^b_\alpha(\mathrm{d}q)_b\right)Y_{\textsf{X}}^\alpha\right]\varepsilon Z_{\textsf{X}}^\perp\left[\frac{p}{\sqrt{\gamma}}\varepsilon n_{\beta}+e^a_\beta(\mathrm{d}q)_a-\varepsilon b^2q\nu_{\beta} \right]\nu^\beta\frac{\mathrm{vol}_{\partial\gamma}}{|\nu^\top|}\,;
\end{equation}
be zero for every $Y\in\mathfrak{X}(\mathcal{M}_1^J)$.

\subsection{Dirichlet boundary conditions}

As we mentioned at the beginning of the paper, the way to deal with homogeneous Dirichlet and Robin boundary conditions differs in some important details. Dirichlet boundary conditions are enforced by restricting the configuration space to those scalar fields which vanish at the boundary. As a consequence of this, the $q$ component $Y_{\textsf{q}}$ of the admissible vector fields $Y\in \mathfrak{X}(\mathcal{M}_1^D)$ must also vanish at the boundary, in particular the $q$ component \eqref{ZQ} of the Hamiltonian vector field must be zero.  By continuity this leads to
\begin{equation}
g\left(Z_{\textsf{X}},\frac{p}{\sqrt{\gamma_X}}\varepsilon n_X +\tau_X.\mathrm{grad}_{\gamma_X} q\right)\Big|\partial\Sigma=Z^\alpha_{\textsf{X}}\left(\frac{p}{\sqrt{\gamma_X}}\varepsilon n_{X\, \alpha} +(e_X)^a_\alpha (\mathrm{d}q)_a \right)\Big|\partial\Sigma=0\,.
\label{condbdryD1}
\end{equation}
We can check now that this expression implies that the boundary integral term \eqref{eq last boundary integral} vanishes and, hence, there are no extra conditions on the vector field. Indeed, first notice that as $Z_{\textsf{X}}|\partial \Sigma\in\Gamma^\partial(X^*T\partial_\Sigma M)$, whenever $Z^\alpha_{\textsf{X}}$ is not zero at the boundary $\partial\Sigma$, equation \eqref{condbdryD1} implies
\begin{equation}\label{equation dphi prop to nu}
\frac{p}{\sqrt{\gamma_X}}\varepsilon n_X+\tau_X.\mathrm{grad}_{\gamma_X}q\varpropto \nu_X\,,
\end{equation}
 and then, the factor
\begin{equation}
Y_{\textsf{q}}-\left(\frac{p}{\sqrt{\gamma_X}}\varepsilon n_{X\, \alpha}+(e_X)^b_\alpha(\mathrm{d}q)_b\right)Y_{\textsf{X}}^\alpha
\end{equation}
of \eqref{eq last boundary integral} is zero, because $Y_{\textsf{q}}|\partial\Sigma=0$ and $g(Y_{\textsf{X}},\nu)|\partial \Sigma=0$ as a consequence of \eqref{nuperp}.

It is useful to rewrite equation \eqref{equation dphi prop to nu} in a different way. If we denote $z_{\textsf{X}}:=Z_{\textsf{X}}|\partial \Sigma =z_{\textsf{X}}^\perp n+\tau.z_{\textsf{X}}^\top$, the condition $Z_{\textsf{X}}|\partial \Sigma\in\Gamma^\partial(X^*T\partial_\Sigma M)$ becomes
 \begin{equation}
 \varepsilon \, \nu_X^\perp z_{\textsf{X}}^\perp + \gamma_X(\nu_X^\top , z_{\textsf{X}}^\top) =0\,,\label{condbdryD2}
\end{equation}
while equation \eqref{condbdryD1} is equivalent to
\begin{align}
\frac{p}{\sqrt{\gamma_X}}z_{\textsf{X}}^\perp+\gamma_X(\mathrm{grad}_{\gamma_X} q, z_{\textsf{X}}^\top)&=0\quad \text{ at }\partial\Sigma\,.\label{condbdryD1-1}
\end{align}

Equations \eqref{condbdryD2} and \eqref{condbdryD1-1} define an homogeneous linear system for the boundary values $z_{\textsf{X}}^\perp,z_{\textsf{X}}^\top$ that has to be solved in terms of $p$ and $\mathrm{d}q$. We always have $z_{\textsf{X}}^\perp=0$ and $z_{\textsf{X}}^{\top}=0$ as solutions. Pointwise, a non-trivial solution is only possible if
\begin{equation}
\left(\varepsilon  \nu_X^\perp \mathrm{grad}_{\gamma_X} q -\frac{p}{\sqrt{\gamma_X}}\nu_X^\top\right)\Big|\partial\Sigma
\label{constraint_surface}
\end{equation}
is zero at $s\in\partial\Sigma$. These conditions are not constraints in the standard sense but rather define sectors in the primary constraint manifold.  The best setting to understand the appearance of these sectors is the study of the parametrized electromagnetic field (see \cite{Nos2}) where the all-important Gauss law appears \textit{precisely} in the same way. Labelling the points of $\mathcal{M}_1^J$ by the support of \eqref{constraint_surface}, we characterize the relevant sectors where the ``number and type'' of the independent components of $z_{\textsf{X}}$ changes. A dinamically relevant sector is the one with elements $(q,p,X)$ such that the associated expression given by \eqref{constraint_surface} has empty support (or equivalently the condition \eqref{equation dphi prop to nu} holds everywhere). In that case $z_{\textsf{X}}$ can be different from zero, and so the dynamics of the parametrized system allows the embeddings ``to advance in time'' and generate a genuine (local) spacetime foliation. If, instead, the $z_{\textsf{X}}$ were zero the integral curve of embeddings would be forced to be ``stuck'' to the same section $X_0(\Sigma)$ of $\partial_\Sigma M$ and, hence, would not be suitable to describe field dynamics in the usual way.

A consistency condition must be imposed now: the Hamiltonian evolution defined by the Hamiltonian vector fields that we have obtained must be compatible with condition \eqref{equation dphi prop to nu} (equivalently the vanishing of \eqref{constraint_surface}), which is a requirement to have proper dynamics for the embeddings as we have just seen. This means that the Hamiltonian vector fields must be tangent to the submanifold formed by the elements $(q,p,X)$ for which \eqref{equation dphi prop to nu} holds. This points out to the existence of additional requirements necessary to have consistent dynamics. This situation exactly mimics the one found in the Hamiltonian treatment of the scalar field with Dirichlet boundary conditions (see \cite{Nos1}) and is in perfect agreement with the known results for the scalar field in the smooth case \cite{Brezis}. From here on, the determination of the infinite chain of conditions necessary to have well defined dynamics for smooth field and embeddings, follows exactly the steps of the GNH algorithm as the main geometric issue involved is the tangency of the Hamiltonian vector fields to the submanifolds defined by the successive conditions. Although they can be obtained in a straightforward manner using Table \ref{table.variations} of appendix \ref{appendix2}, they are somewhat complicated and their particular form is not specially illuminating, hence we do not give them here.

As a final remark notice that if $\nu_X^\perp=0$ at $X(\partial\Sigma)$, or equivalently $\nu_X^\alpha$ is tangent to $X(\Sigma)$ at $X(\partial\Sigma)$, the condition \eqref{constraint_surface} immediately implies $p|{\partial\Sigma}=0$ (as $\nu^\top$ never vanishes). This condition is found in the Hamiltonian treatment of the scalar field with Dirichlet boundary conditions \cite{Nos1}, where a special foliation with $z_{\textsf{X}}^\perp=1$ and $z_{\textsf{X}}^\top=0$ is used.

\subsection{Robin boundary conditions}
In order to implement the Robin boundary conditions we allow the fields $q$ to take non-zero values at $\partial\Sigma$  and consider $b$ arbitrary (if it is zero we would be in the Neumann case). In general the boundary integral \eqref{eq last boundary integral} would not vanish as before, so according to equation \eqref{eqHamVfields} we have to require it to be zero for every $Y\in\mathfrak{X}(\mathcal{M}_1^R)$. Denoting again $z_{\textsf{X}}:=Z_{\textsf{X}}|\partial\Sigma$, we have
\begin{equation}
  z_{\textsf{X}}^\perp\left[\frac{p}{\sqrt{\gamma_X}}\varepsilon n_{X\, \beta}+(e_X)^a_\beta(\mathrm{d}q)_a-\varepsilon b_X^2q\nu_{X\, \beta} \right]\nu_X^\beta\Big|\partial\Sigma=0\label{condbdryR1}
\end{equation}
which plays a role analogous to the one of equation \eqref{condbdryD1} in the Dirichlet case, despite their very different origin.

Wherever $z_{\textsf{X}}^\perp\neq0$, equation \eqref{condbdryR1} implies
\begin{equation}
  \frac{p}{\sqrt{\gamma_X}}\varepsilon n_X+\tau_X.\mathrm{grad}_{\gamma_X} q-\varepsilon b_X^2q\nu_X\in T_{X\circ \imath_\partial}\partial_\Sigma M\,.\label{condbdryR1-1}
\end{equation}
Condition \eqref{condbdryR1-1} is the analogous of \eqref{equation dphi prop to nu}. It is interesting to mention that in the Dirichlet case we have that
\[
\frac{p}{\sqrt{\gamma_X}}\varepsilon n_X+\tau_X.\mathrm{grad}_{\gamma_X} q
\]
is normal to $\partial_\Sigma M$, in the Neumann case ($b=0$) it is tangent to it, and in the Robin it is neither normal nor tangent. Clearly we can recover the Neumann case from the Robin one by making $b=0$, however, there is no such way to pass from Robin to Dirichlet. This shows once more the intrinsically different nature of these boundary conditions.

\bigskip

As in the Dirichlet case, we can reexpress \eqref{condbdryR1-1} in a language based entirely on objects defined on $\Sigma$.  We have the equations
\begin{align}
z_{\textsf{X}}^\perp\left[\frac{p}{\sqrt{\gamma_X}}\nu_X^\perp+\mathrm{d}q(\nu_X^\top)-\varepsilon b_X^2q \right]\Big|\partial\Sigma=0\\
 \varepsilon \, \nu_X^\perp z_{\textsf{X}}^\perp + \gamma_X(\nu_X^\top, z_{\textsf{X}}^\top) =0.
\end{align}
Again, these should be considered as a system of equations for $z_X^\perp,z_X^\top$, which has non-trivial solutions when the following compatibility requirement holds
\begin{equation}
\mathrm{d}q(\nu_X^\top)= \varepsilon  b_X^2 q-\frac{p}{\sqrt{\gamma_X}}\nu^\perp_X  \quad\textrm{ on }\partial\Sigma\,.
\label{bdry-constr-Robin}
\end{equation}
Notice that the vector field
\[
\xi_X:=\frac{\nu^\top_X}{\sqrt{\gamma_X(\nu^\top_X,\nu^\top_X)}}
\]
is the unit outer normal (according to $\gamma_X$) to $\partial\Sigma$ and, hence, $$\mathrm{d}q(\nu_X^\top)= \sqrt{\gamma_X(\nu^\top_X,\nu^\top_X)} \, \mathrm{d}q(\xi_X)$$ is proportional to the normal derivative of the scalar field $q$.

\bigskip

We can see again the same phenomenon that we found for the Dirichlet boundary conditions. There are sectors in the primary constraint hypersurface $\mathcal{M}_1^R$ where the components $z_{\textsf{X}}$ must necessarily vanish whereas in other parts of $\mathcal{M}_1^R$, formed by elements $(q,p,X)$ for which \eqref{condbdryR1-1}/\eqref{bdry-constr-Robin} hold, the $z_{\textsf{X}}$ components may be different from zero.

The situation at this point is conceptually equivalent to the one that we explained in the Dirichlet case: the conditions \eqref{condbdryR1-1}/\eqref{bdry-constr-Robin} behave as secondary constraints whose stability under the dynamics given by the Hamiltonian vector fields must be enforced. From a geometric point of view this is a tangency requirement that provides an infinite chain of constraints (whose relatives in the unparametrized case are explicitly given in \cite{Nos1}).

\bigskip

We can make the same final remark as in the Dirichlet case, if we consider embeddings $X$ such that the normal to the boundary $\nu_X$ and the normal $n_X$ to $X(\Sigma)$ are orthogonal (i.e.\ such that $\nu^\perp_X=0$), and taking into account that, in this case, $\nu_X^\top=\xi_X$, we recover the condition
\begin{equation}
\Big(\mathrm{d}q(\xi_X) -\varepsilon   b_X^2 q \Big)|\partial\Sigma=0
\end{equation}
 that we found for the non-parametrized case in \cite{Nos1}.

%
%
\section{Comments and conclusions}\label{sec_conclusions}

The main result of the paper is the precise description of the Hamiltonian formulation for a parametrized scalar field defined in a bounded spatial region with or without boundaries. When no boundaries are present we recover the results of \cite{Hajicek}. We have considered boundary conditions of the Dirichlet and Robin types (Neumann boundary conditions are a particular example of the latter) and worked within the class of smooth fields and spacelike embeddings. We have obtained the constraint submanifold in phase space and the concrete form of the Hamiltonian vector fields. Contrary to the statements in \cite{Andrade}, we have found no obstructions to get the full description in the Robin case. We expect that our detailed formulation can be used to extend the polymer quantization of the scalar field to spatial manifold with boundaries, in particular in the 1+1 dimensional case. Other systems that can be considered from this perspective are Einstein-Rosen waves coupled with massless scalar fields \cite{BGV}

\bigskip

There are a number of interesting facts that we would like to mention. First, the solutions to equation $i_Z\omega=0$, giving the Hamiltonian vector fields, can be obtained pointwise on the primary constraint hypersurface, however, it is a non-trivial issue to understand to what extent they define smooth vector fields on $\mathcal{M}_1^J$ or a submanifold thereof. There are a number of regularity issues that must be characterized and understood. For instance, there are sectors in the primary constraint hypersurface where some components of the vector fields are forced to be zero whereas in others, defined by the vanishing of specific functions of the configuration variables and momenta, they can be different from zero. We have paid particular attention to a sector which is specially important from the dynamical point of view: the one defined by conditions \eqref{equation dphi prop to nu} and \eqref{condbdryR1-1} for the Dirichlet and Robin cases respectively.

Second, for parametrized models the energy is zero and, hence, the appearance of different sectors is even more important in the Lagrangian symplectic approach. In this approach the canonical symplectic structure is pulled back to the tangent bundle of the configuration space with the help of the fibre derivative, and the dynamics is obtained by finding the Hamiltonian vector fields given by its degenerate directions \cite{Gotay}. The standard Hamiltonian approach can be mirrored in this setting, where no constraints show up and all the subtleties associated with the singularity of the system must manifest themselves under the guise of regularity issues of a nature similar  to the one of those discussed in this paper.

Finally, we would like to point out that the Hamiltonian description accommodates, in a straightforward way, some non stationary situations where the metric $g_\partial$ induced on the boundary $\partial_\Sigma M$ has no timelike Killing vector fields. In the Robin case, it is furthermore possible to introduce non-constant boundary conditions encoded on the value of the scalar function $B$ on $\partial_\Sigma M$. These non-stationary situations fit quite naturally in our approach as stationarity plays no role in any part of it but, of course, the integrability of the Hamiltonian vector fields must be considered with due care.

\bigskip

The method that we have used in the paper is quite general and can be applied in a similar way to other parametrized field theories  (gauge fields, gravity, etc.) defined in different types of spatial regions (bounded and/or unbounded). We plan to explore them in the near future.

%
%
\section*{Acknowledgments}

This work has been supported by the Spanish MINECO research grants FIS2012-34379, FIS2014-57387-C3-3-P and the  Consolider-Ingenio 2010 Program CPAN (CSD 2007-00042). Juan Margalef-Bentabol is supported by a ``la Caixa'' fellowship.

%
%
\begin{appendices}

\section{Variations in \texorpdfstring{$\bm{\textrm{Emb}(\Sigma,M)}$}{Emb}}\label{appendix2}

A detailed description of the infinite dimensional manifold $\mathrm{Emb}(\Sigma,M)$ of embeddings  $X:\Sigma\hookrightarrow M$ in the case where $(M,g)$ is a Riemannian  manifold can be found in \cite{Bauer}. References \cite{Isham1,Hajicek,Kuchar1}  provide a summary on the relevant results concerning the manifold $\mathrm{Emb}_{g\textrm{-sl}}(\Sigma,M)$.  We refer the reader to those papers and references therein for further details. In this appendix we just list a number of expressions for the variations of several geometric tensor fields on $\mathrm{Emb}(\Sigma,M)$ that are necessary to get the Hamiltonian formulation for the systems that we consider here. We give them in Table \ref{table.variations}, where we have corrected the typos in equations (9.8) and (9.10) appearing in section 9 of \cite{Kuchar1}.

\begin{table}
\[
\begin{array}{|l|l|}
  \hline
  \multicolumn{1}{|c|}{f_X} & \multicolumn{1}{|c|}{D_{(X,\dot{X})}f} \\ \hline\hline
  b_X:=B\circ X & \rule{0ex}{2.8ex}D_{(X,\dot{X})}b= \mathrm{d}B(\dot{X}) \\
  \nu_X:=\mathcal{V}\circ X & D_{(X,\dot{X})}\nu=\nabla_{\dot{X}} \mathcal{V} \\ \hline
  (e_X)^a_\alpha  & \rule{0ex}{2.8ex}\displaystyle D_{(X,\dot{X})}e^a_\alpha=\left(K^a_b\dot{X}^{\top b}+\varepsilon\gamma^{ab}(\mathrm{d}\dot{X}^\perp)_b\right)\varepsilon n_\alpha-e^b_\alpha \left(\nabla_b \dot{X}^{\top a}-\dot{X}^\perp K_b^a\right) \\
  (\tau_X)_a^\alpha  & \rule{0ex}{2.8ex}\displaystyle D_{(X,\dot{X})}\tau_a^\alpha=\left(K_{ab}\dot{X}^{\top b}+\varepsilon(\mathrm{d}\dot{X}^\perp)_a\right)\varepsilon n^\alpha+\tau^\alpha_b\left(\nabla_a \dot{X}^{\top b}-\dot{X}^\perp K^b_a\right)\\ \hline
  n_X^\alpha   & \displaystyle D_{(X,\dot{X})}n^\alpha=-\tau^\alpha_a\Big(K^a_b\dot{X}^{\top b}+\varepsilon \gamma^{ab}(\mathrm{d}\dot{X}^\perp)_b\Big)\\
  (n_X)_\alpha:=g_{\alpha\beta}n_X^\beta & \displaystyle D_{(X,\dot{X})}n_\alpha=-e_\alpha^a\Big(K_{ab}\dot{X}^{\top b}+\varepsilon (\mathrm{d}\dot{X}^\perp)_a \Big)
  \\ \hline
  (\gamma_{X})_{ab} & \rule{0ex}{2.8ex}\displaystyle D_{(X,\dot{X})}\gamma_{ab}=\gamma_{bc}\nabla_a \dot{X}^{\top c}+\gamma_{ac}\nabla_b\dot{X}^{\top c}-2\dot{X}^\perp K_{ab}\\
  \gamma_X^{ab} & \rule{0ex}{2.8ex}\displaystyle D_{(X,\dot{X})}\gamma^{ab}=-\nabla^a\dot{X}^{\top b}-\nabla^b\dot{X}^{\top a}+2\dot{X}^\perp K^{ab}
  \\ \hline
  \mathrm{vol}_{\gamma_X} & \rule{0ex}{2.8ex}\displaystyle D_{(X,\dot{X})}\mathrm{vol}_{\gamma}=\Big(\nabla_a\dot{X}^{\top a}-\dot{X}^\perp K_a^a\Big) \mathrm{vol}_{\gamma}\\
  \sqrt{\gamma_X} & \displaystyle D_{(X,\dot{X})}\sqrt{\gamma}=\Big(\nabla_a\dot{X}^{\top a}-\dot{X}^\perp K_a^a\Big)\sqrt{\gamma}\\ \hline
\end{array}
\]
\caption{\label{table.variations}
Variations of the relevant geometric tensor fields on $\mathrm{Emb}(\Sigma,M)$. Here and in the following, $K_a^b$ denotes the Weingarten map associated with $X(\Sigma)\subset M$, with indices lowered and raised with the help of $\gamma_{ab}$ and $\gamma^{ab}$, and $\nabla$ denotes the Levi-Civita connection of $(\Sigma,\gamma)$.}
\end{table}

\vfill\eject

Using Table \ref{table.variations},  it is immediate to obtain $Y_{\textsf{P}\alpha}=D\left(-\varepsilon n_{\alpha}\mathcal{H}_\perp -e^b_{\alpha}\mathcal{H}_b\right)$ in the form (dropping the $X$ subindex)

\begin{align}
\frac{(Y_\textsf{P})_\alpha}{\sqrt{\gamma}}&=-\left(\frac{p}{\sqrt{\gamma}}\varepsilon n_\alpha+e^b_\alpha(\mathrm{d}q)_b\right)Y_{\textsf{p}}+\left(\gamma^{ac}(\mathrm{d}q)_c \varepsilon n_\alpha-e^a_\alpha \frac{p}{\sqrt{\gamma}}\right)(\mathrm{d}Y_{\textsf{q}})_a-\varepsilon n_\alpha m^2q Y_{\textsf{q}}\nonumber\\
            &\phantom{=}+ \left[-\frac{p}{\sqrt{\gamma}}(\mathrm{d}q)_b\left(K^b_a Y_\textsf{X}^{\top a}+\varepsilon\gamma^{ab}(\mathrm{d}Y_\textsf{X}^\perp)_a\right)-\varepsilon \left(\nabla^aY_\textsf{X}^{\top b}-Y_\textsf{X}^\perp K^{ab}\right)(\mathrm{d}q)_a(\mathrm{d}q)_b\right.\nonumber\\
            &\phantom{=+\varepsilon n_\alpha[}-\left.\left(\frac{\mathcal{H}_\perp}{\sqrt{\gamma}}-\frac{p^2}{\gamma}\right)\left(\nabla_aY_\textsf{X}^{\top a}-Y_\textsf{X}^\perp K_a^a\right)\right]\varepsilon n_\alpha\label{up}\\
            &\phantom{=}+e^a_\alpha\left[ \left(\nabla_a Y_\textsf{X}^{\top b}-Y_\textsf{X}^\perp K_a^b\right)\frac{p}{\sqrt{\gamma}}(\mathrm{d}q)_b+\varepsilon \frac{\mathcal{H}_\perp}{\sqrt{\gamma}}\left(K_{ab}Y_\textsf{X}^{\top b}+\varepsilon(\mathrm{d}Y_\textsf{X}^\perp)_a\right)\right]\nonumber\,.
\end{align}

Substituting the ambient spaces $M$ and $\Sigma$ by $\partial_\Sigma M$ and $\partial \Sigma$, and considering the induced embeddings $X^\partial:\partial\Sigma\hookrightarrow\partial_\Sigma M$ we can obtain a similar table of variations, where now the relevant  objects are associated with the boundary $\partial\Sigma$. For instance we have to use the Weingarten map $K_{\phantom{\partial}a}^{\partial b}$ of $X(\partial\Sigma)\subset \partial_\Sigma M$ and the Levi-Civita connection $\nabla^\partial$ of $(\partial\Sigma,\gamma_{\partial})$. Also, for any given $W_X\in\Gamma^\partial(X^*TM)$, we have to use the decomposition
\begin{equation}
  W_X|\partial\Sigma=W_X^{\partial\perp}\theta_X+\tau^\partial_X W_X^{\partial\top}
\end{equation}
where $\tau_X^\partial:= TX^\partial$, $\varepsilon W_X^{\partial\perp}:=g(W_X|\partial\Sigma,\theta_X)$ and $W_X^{\partial\top}\in\mathfrak{X}(\partial\Sigma)$ is defined by $\tau_X^\partial.W_X^{\partial\top}:=W_X-W_X^{\partial\perp}\theta_X$. With all these elements we find that $Y^\partial_{\textsf{P}\alpha}=D\left(-\varepsilon \theta_{\alpha} \mathcal{H}^\partial_\perp \right)$ can be expressed as (dropping again the $X$ subindex)
\begin{align}
\frac{(Y^\partial_\textsf{P})_\alpha}{\sqrt{\gamma_{\partial}}}&=\left[(\nabla^\partial_a Y_\textsf{X}^{\partial\top a}-Y_\textsf{X}^{\partial\perp}K^{\partial a}_{\phantom{\partial}a})\frac{b^2}{2} q^2 +b\left(Y_\textsf{X}^{\partial\perp}\theta^\beta(\mathrm{d}b)_\beta+Y_\textsf{X}^{\partial\top\beta}(\mathrm{d} b)_\beta\right)q^2+b^2q Y_\textsf{q}\right]\varepsilon\theta_\alpha\nonumber\\
&\phantom{=}-\frac{b^2}{2}q^2\varepsilon (e^\partial)^a_\alpha\Big[K_{ab}^\partial Y_\textsf{X}^{\partial \top b}+\varepsilon(\mathrm{d}Y_\textsf{X}^\perp)_a\Big]\label{down}\,.
\end{align}

 As we have to make contact with the decomposition $W=W^\perp n+\tau.W^\top$ associated with $n$ and $\tau$, we need to use the following relations
\begin{equation}
  W^{\partial\perp}=\frac{W^\perp}{|\nu^\top|}\,,\qquad\qquad W^{\partial\top}=W^\top+\frac{\varepsilon\nu^\perp W^\perp}{|\nu^\top|^2}\nu^\top\,.\label{eq:thetan}
\end{equation}
Plugging \eqref{eq:thetan} in \eqref{down} and using \eqref{up}, we obtain the final form of \eqref{granecuacion1}.

An interesting application of the results of Table \ref{table.variations} is the derivation of the hypersurface deformation algebra \cite{Kuchar1}. Considering $V=V^\perp n + \tau. V^\top$ and $W=W^\perp n + \tau. W^\top$, vector fields on the space of embeddings, we can define
\begin{equation}
[V,W]:= D_{V} W- D_{W}V\nonumber
\end{equation}
or, equivalently, $[V,W]_X:= D_{(X,V_X)} W- D_{(X,W_X)} V$. From these, it is immediate to obtain the decomposition
\begin{align}
[V,W] &=\Big(D_V W^\perp - D_W V^\perp+  \mathrm{d} V^\perp (W^\top)-  \mathrm{d} W^\perp(V^\top) \Big) n\nonumber\\
&\phantom{=}+ \tau.\Big(D_{V}W^\top-D_{W}V^\top+\varepsilon (V^\perp \textrm{grad}_{\gamma_X}  W^\perp- W^\perp \textrm{grad}_{\gamma_X} V^\perp) -[V^\top,W^\top]\Big)\,.\nonumber
\end{align}
\end{appendices}


\begin{thebibliography}{99}

\bibitem{Dirac1} P. A. M. Dirac, Can. J. Math. \textbf{3} (1951) 1.

\bibitem{Kuchar3} K. V. Kucha\v{r}, J. Math. Phys. \textbf{17} (1976) 801.

\bibitem{Hajicek} P. H\'aj\'{\i}\v{c}ek and C. Isham, J. Math. Phys. \textbf{37} (1996) 3505.

\bibitem{Isham1} C. J. Isham and K. V. Kucha\v{r}, Ann. Phys. \textbf{164} (1985) 128.

\bibitem{Torre} C. G. Torre, J. Math. Phys. \textbf{33} (1992) 3802.

\bibitem{LV1} A. Laddha and M. Varadarajan, Phys. Rev. D\textbf{83} (2011) 025019.

\bibitem{LV2} A. Laddha and M. Varadarajan, Class. Quant. Grav. \textbf{27} (2010) 175010.

\bibitem{Andrade} T. Andrade, D. Marolf and C. Deffayet, Class. Quantum Grav. \textbf{28} (2011) 105002.

\bibitem{Kuchar} K. V Kucha\v{r}, Phys. Rev. D\textbf{4} (1971) 955.

\bibitem{BGV} J. F. Barbero G., I. Garay and E. J. S. Villase\~nor, Phys. Rev.  Lett. \textbf{95} (2005) 051301.

\bibitem{LR} J. F. Barbero G. and E. J. S. Villase\~nor,  Living Rev. Rel. \textbf{13} (2010) 6.

\bibitem{Kuchar1} K. V. Kucha\v{r} J. Math. Phys, \textbf{17} (1976) 777.

\bibitem{Kuchar2} K. V. Kucha\v{r} J. Math. Phys, \textbf{17} (1976) 792.

\bibitem{Dirac2} P. A. M. Dirac, \textit{Lectures on Quantum Mechanics} (Dover Publications Inc., 2001).

\bibitem{GN}  M. Gotay and J. Nester, \textit{Generalized constraint algorithm and special presymplectic manifolds}
(Lecture Notes in Mathematics 775, Geometric Methods in Mathematical Physics), Springer Berlin Heidelberg (1980).

\bibitem{Gotay} M. J. Gotay, \textit{Presymplectic Manifolds, Geometric Constraint Theory and the Dirac-Bergmann
Theory of Constraints}, Thesis, Center for Theoretical Physics of the University of Maryland (1979).

\bibitem{GNH} M. Gotay, J. Nester and G. Hinds, J. Math. Phys. \textbf{19} (1978) 2388.

\bibitem{Anco1} S. C. Anco and R. S. Tung, J. Math. Phys. \textbf{43} (2002) 3984.

\bibitem{Anco2} S. C. Anco and R. S. Tung, J. Math. Phys. \textbf{43} (2002) 5531.

\bibitem{Nos2} J. F. Barbero G., J. Margalef-Bentabol and E. J. S. Villase\~nor, 	arXiv:1511.00826 [hep-th].

\bibitem{Nos1} J. F. Barbero G., J. Prieto and E. J. S. Villase\~nor, Class. Quant. Grav. \textbf{31} (2014) 045021.

\bibitem{Brezis} H. Brezis, \textit{Functional Analysis, Sobolev Spaces and Partial Differential Equations} (Springer-Verlag, 2011).

\bibitem{Michor2} A. Kriegl and P. W. Michor. \textit{The convenient setting of global analysis}, volume 53 of Mathematical Surveys and Monographs. American Mathematical Society, Providence, RI, (1997).

\bibitem{Bauer} M. Bauer, P. Harms and P. W. Michor, SIAM J. Imaging Sci. \textbf{5} (2012) 244.

\end{thebibliography}
\end{document}